# The Dynamics of Centaurs in the Vicinity of the 2:1 Mean Motion Resonance of Neptune and Uranus Trojan Region

Jeremy Wood[1,2] Jonathan Horner[2,3]

*[1] Hazard Community and Technical College 1 Community College Drive Hazard, Ky USA 41701*
*[2] University of Southern Queensland West St, Toowoomba QLD 4350, Australia*
*[3] Australian Centre for Astrobiology, UNSW Australia, Sydney, NSW 2052, Australia*

**Summary:** In this work we present the results of a suite of dynamical simulations following the orbital evolution of 8,022 hypothetical Centaur objects. These Centaurs begin our integrations on orbits in the vicinity of the 2:1 mean motion resonance with Neptune, and we follow their dynamical evolution for a period of 3 Myr under the gravitational influence of a motionless Sun and the four Jovian planets. The great majority of the test particles studied rapidly escaped from the vicinity of the 2:1 mean motion resonance of Neptune and diffused throughout the Solar System. The average libration time of Centaurs in the vicinity of 2:1 mean motion resonance of Neptune was found to be just 27 kyr. Although two particles did remain near the resonance for more than 1 Myr. Upon leaving the vicinity of the 2:1 resonance, the majority of test particles evolved by a process of random walk in semi-major axis, due to repeated close encounters with the giant planets.

## Introduction

Small bodies in the Solar system exist in orbits that extend from within that of the Earth to beyond Neptune's, and are collectively the detritus left behind from the formation of our planetary system. The hazard that these small bodies pose to Earth has been brought to public attention in recent years by the meteor strike in Chelyabinsk, Russia, in Feb. 2013 that injured over 1,100 people [1]. Numerous very close approaches between small bodies and the planets have also been observed, such as the near miss of 2012 DA14 [2][3] and 2014 HQ124 [4] with Earth and the close approach of comet C/2013 A1 (Siding Spring) [5][6][7] to Mars in Oct. 2014.

Two spectacular collisions between a planet and a small body occurred in the 20$^{th}$ century. In 1908 a small body exploded over Tunguska, Russia flattening an area of 2,150 km$^2$ [8][9][10] and in 1994, comet D/1993 F2 (Shoemaker-Levy 9) split into 21 pieces before colliding with Jupiter [11]. The collision left dark spots on the planet which lasted over a year [12]. The size and persistence of the scars left behind by the impacting cometary fragments illustrates the damage collisions between small bodies and planets can cause. Indeed, the energy contained in an impacting small body can be enough to cause an extinction level event. Such collisions have been proposed in order to explain a number of historical mass extinction events – most famously that which occurred some 65 Myr ago, resulting in the death of the dinosaurs [13]. Thus, one reason for studying small bodies of the Solar system is because of the potential hazard they pose.

Small bodies with initial orbits near Earth are not the only ones with the potential for an Earth strike. The immediate threat to Earth comes from three distinct populations of objects – the Near-Earth Objects (such as the Chelyabinsk impactor), the short-period comets (such as comet Shoemaker-Levy 9), and the long-period comets (such as comet Siding Spring). Each of these threatening populations are dynamically unstable, and are continually replenished from reservoirs of rocky or icy bodies further from the Sun [14][15][16][17]. The exact definition of each class of small body differs from one researcher to the next. Traditionally, Near Earth Objects (or NEOs) have been classified into three groups [18]. These are:

Atens – Objects with orbital semi-major axes less than that of the Earth whose orbits cross our own
Apollos – Objects with orbital semi-major axes greater than that of the Earth's, whose orbits cross the Earth's
Amors – Objects with orbits exterior to Earth's, and perihelia within 1.3 AU, that do not cross Earth's orbit

More recently, the discovery of Near-Earth Asteroids whose orbits are entirely within that of the Earth has resulted in the addition of a fourth sub-class – the Atira or Apohele asteroids, whose aphelia are closer to the Sun than the Earth's perihelion distance (e.g. [19]). In addition, those NEOs that can approach to within 0.05 AU of Earth are classified as Potentially Hazardous Astereoids or PHAs [20][21]. Currently 1,548 PHAs are known [20].

Objects that display noticeable activity (such as outgassing, or the presence of a tail) are classified as cometary. Amongst those bodies, the short-period comets are defined to be those having an orbital period of less than 200 years [22]. The Minor Planet Center refers to these as periodic comets [23]. The population of short period comets is replenished by the Centaurs [24]. The Centaurs are a group of small rock-ice bodies with semi-major axes between Jupiter and Neptune and perihelia beyond that of Jupiter's [21][22][25]. The frequency of the flux of Centaurs into a new Earth-crossing orbit is estimated to be 1 every 880 years [26]. Though the possibility of an eventual Earth collision with a Centaur is an order of magnitude lower than that of a Main Belt Asteroid, the threat is non-zero and more likely than a collision with a Long Period Comet [14][15][17].

Numerical studies show that small bodies in orbits beyond Earth's can be transported to the inner Solar system over time via the gravitational perturbation of their orbits [27]. Over their lifetime, small bodies may also enter into a mean motion orbital resonance with a planet. A mean motion orbital resonance occurs when the orbital periods of two bodies exist in a ratio of two small integers $r:s$. For example, Saturn orbits the Sun in about 29.5 years and Jupiter in about 11.86 years.

Dividing these orbital periods yields almost 2.5 which is equal to the ratio 5:2. Thus, it can be said that Saturn and Jupiter are nearly in a 5 to 2 mean motion orbital resonance. In other words, Jupiter completes five complete orbits in the time it takes Saturn to complete two. For a body to be considered stuck in a resonance, the principal resonant angle associated with that resonance must also librate or change very slowly [40]. In the case of the 2:1 mean motion resonance of Neptune, the principal resonant angle is defined by

$$2\lambda_N - \lambda - \varpi \tag{1}$$

Where $\lambda_N$ is the mean longitude of Neptune, $\lambda$ is the mean longitude of the small body ($\lambda = M + \varpi$, where $M$ is the mean anomaly) and $\varpi$ is the longitude of perihelion of the small body.

Small bodies in mean motion resonances can either be in stable or unstable orbits. Some resonances are unstable due to the fact that they overlap with others. Over time, the eccentricity of these orbits increases leading to planetary orbit crossings or the objects colliding with the Sun [28][29]. Small bodies remain in these resonances for relatively short amounts of time. However, populations of small bodies also exist in stable mean motion resonances where they can remain for relatively longer periods of time. Examples include Jupiter Trojans (1:1 resonance with Jupiter; e.g. [30][31]) Neptune Trojans (1:1 resonance with Neptune; e.g. [32]), Hilda asteroids (3:2 resonance with Jupiter; e.g. [33]) and Plutinos (2:3 resonance with Neptune; e.g. [34]).

When a small body enters a stable mean motion orbital resonance with a planet, it becomes stuck in the resonance meaning that the semi-major axis of its orbit librates (oscillates in a quasi-periodic fashion) about the resonance location. This is known as resonance sticking [29][35].

Small bodies which orbit between Mars and Jupiter (in the asteroid belt) have received more attention in the literature than those that orbit between Jupiter and Neptune (the Centaurs) [36]. In this work, we present the initial results of a study of the dynamical behavior of bodies trapped in the 2:1 mean motion resonance of Neptune, and their orbital evolution after leaving it. First, in the following section, the theory of mean motion resonances, taxonomy of small bodies, and dynamical classification of Centaurs are discussed. We then present the method used to carry out the study, and describe how the resulting data were analyzed. Finally, we present and discuss our results, and detail our plans for future work.

## Theory

The location of a given mean motion resonance for a planet, $a_{MMR}$, can be derived using the semi-major axis, $a_p$, of the planet, and the orbital period ratio of the resonance in question (*s/r*), using Kepler's 3<sup>rd</sup> Law. The result is:

$$a_{MMR} = a_p \left(\frac{s}{r}\right)^{2/3} \qquad (2)$$

[37]. Here $a_p$ is a planet's semi-major axis. The resonance is exterior to the planet's orbit if $r < s$ and interior if $r > s$. If $r = s$ then any small body in that resonance is called a Trojan asteroid [37][38]. In this paper, interior resonances are stated using the following format: "larger number:smaller number". For example the 2:1 mean motion resonance of Neptune is interior to the planet Neptune, but the 1:2 resonance is exterior to Neptune. Resonances actually exist in a region of *a-e* space, rather than at just one location, so it is possible that two resonances can overlap [39]. The location of the 2:1 mean motion resonance of Neptune is centered at 18.94 AU from the Sun. This is very close to the orbit of the planet Uranus, which lies at $a$ = 19.19 AU. Neptune and Uranus are therefore very nearly in a 2:1 resonance with each other (with a period ratio of ~1.96), and as a result, the 2:1 resonance of Neptune and the Trojan region of Uranus overlap in *a-e* space. This potentially increases the instability of the orbits of any small bodies librating in this region [37][40].

The gravitational influence of the planets also perturbs the osculating orbital parameters of small bodies, even when the small body is not in a mean motion resonance. This is especially true during close planetary encounters. As bodies are often classified using osculating orbital parameters, the classification of a small body will often change during its dynamical lifetime [24]. Common orbital quantities used to classify small bodies include the semi-major axis, $a$, eccentricity, $e$, perihelion distance, $q$, and the Tisserand parameter, $T_p$, defined as:

$$T_p = \frac{1}{\frac{a}{a_p}} + 2\sqrt{\frac{a}{a_p}(1-e^2)}\cos\Delta i \qquad (3)$$

Where $\Delta i$ is the small body's orbital inclination with respect to the plane of the planet's orbit [41]. At any moment in time, the Tisserand parameter for a given small body will be different for the different planets. The Tisserand parameters of Jupiter, Saturn, Uranus and Neptune can be denoted as $T_J$, $T_S$, $T_U$ and $T_N$ respectively.

Over the years, a variety of different classification schemes have been put forward to identify different types of Solar system small bodies (e.g. [42][22][25]). In this work, we follow the classification scheme detailed in Table 1.

*Table 1 One possible taxonomy for small bodies of the Solar system.*

| TYPE | $T_J$ | $a$ (AU) | $q$ (AU) | REFERENCE |
|---|---|---|---|---|
| **Centaur** | - | $a_{Jupiter} < a < a_{Neptune}$ | $> a_{Jupiter}$ | [21] |
| **Encke-Type Comet** | $> 3$ | $< a_{Jupiter}$ | - | [43] |

| Jupiter Family Comet | $2 < T_J < 3$ | - | - | [43] |
|---|---|---|---|---|
| Halley-Type Comet | $< 2$ | $< 40$ | - | [42] |
| Trans Neptunian Object (TNO) | - | $> a_{\text{Neptune}}$ | - | [21] |
| Ambi-Neptunian Object | | $> a_{\text{Neptune}}$ | $< a_{\text{Neptune}}$ | |
| Kuiper Belt Object (KBO) | - | $a_{\text{Neptune}} < a \leq 48$ | - | [35] |
| Scattered Disk Object (SDO) | - | $48 < a \leq 1{,}000$ | $> a_{\text{Neptune}}$ | [35] |
| Oort Cloud Object | - | $> 1{,}000$ | - | [44] |

As for Centaurs, [41] and [36] state that Centaurs can be categorized in one of two dynamical classes referred to as 'random walk' and 'resonance hopping'. If the standard deviation, $\sigma$, of the semi-major axis values of a Centaur varies in time according to a power law given by:

$$\sigma = (2Dt)^H \tag{4}$$

then the Centaur is classified as a random-walk Centaur. Here $t$ is time, $D$ is the diffusion coefficient and $H$ is the Hurst exponent [45]. In this case the semi-major axis is also said to undergo diffusion. From this equation it can also be shown that

$$log_{10}(\sigma) = H log_{10}(t) + constant \tag{5}$$

i.e. the log of the standard deviation varies linearly with the log of time. Centaurs whose semi-major axis values are not dominated by diffusion tend to jump from one resonance to another. These bodies are classified as resonance-hopping Centaurs. Resonance sticking dominates their dynamics. Centaurs may also display behavior of both types during their lifetimes [41].

## Method

8,022 massless test particles were integrated for 3 Myr in the 6-Body problem (a motionless Sun; Jupiter, Saturn, Uranus, Neptune and test particle) using the regularized mixed-variable symplectic (RMVS) method in the SWIFT software package [46]. The time step used was 40 days, and data were output at intervals of 2 kyr. The initial values for $a$, $e$ and $i$ of the test particle orbits were randomly chosen from these ranges: 18.92 AU $\leq a \leq$ 19.16 AU, $0 < e \leq 0.7$, and $0° \leq i \leq 40°$. The initial $a$, $e$ and $i$ of the orbits of the giant planets were obtained directly from the standard SWIFT installation. The unusual range of $a$ was because of the way in which the SWIFT integrator behaves. An $a$ range symmetric about the resonance location was input, however this is the range that SWIFT used. Though odd, the range is totally usuable for this study. Both the planets and the test particles were initially placed at random locations in their orbits. The arguments of perihelion and longitudes of the ascending node were also randomised. Test particles were removed from the simulation by colliding with a planet, approaching too close to the Sun (~0.005 AU), obtaining a parabolic or hyperbolic orbit ($e \geq 1$), or by entering the Oort Cloud (defined as reaching a barycentric distance of 1,000 AU).

Once the simulations were complete, a random sample of 75 test particles was chosen, and their orbital elements plotted as a function of time, in order to determine approximate maximum allowed amplitudes of osculating and average semi-major axis within which a test particle is considered to display libration-like behavior about the 2:1 Neptunian mean motion resonance. Since the chosen data output period was too large for the true resonant angles to be determined (a compromise chosen to ensure the simulations could run in a reasonable amount of time), test particles are said to display only pseudo-librational behavior about the resonance, rather than true libration.

To determine the libration times, the dynamical lifetime of each test particle is divided into 10 kyr windows of time. If the last window has a duration of < 10 kyr then it is ignored. For example, if the dynamical lifetime of a test particle is 216 kyr then its lifetime is divided into 21 10 kyr windows, and the last 6 kyr are ignored.

To determine the longest consecutive libration time of a test particle, the average semi-major axis in each window is calculated and subtracted from the resonance location to calculate the amplitude. The maximum osculating semi-major axis value in the window is also determined. If both the amplitude of the average and osculating values of *a* are smaller than their respective maximum limits of 0.125 AU and 1.15 AU respectively from the resonance location, then the test particle is said to display libration-like behavior in the window. The largest number of consecutive windows displaying libration-like behavior is then determined, and multiplied by 10 kyr to find the total libration time.

The dynamical class of a sample of test particles is determined using the quantitative method of [41]. This method begins by finding the time interval within which 10 values of osculating semi-major axis occur. The dynamical lifetime is then divided into windows of time each equal to this interval, allowing consecutive windows to overlap by half a time interval. Any windows of duration longer than one-quarter the total time are discarded. The standard deviation of semi-major axis in each time window is calculated, and then the values are averaged over all windows. Following this, the process is repeated using up to 16 different window time intervals. To find the duration of other time intervals, we take the log base ten of the total number of data points during the dynamical lifetime of the test particle and subtract the log base ten of the fewest data points in any window (which is 10) and then dividing the difference by 16. Finally, find the slope of the graph of $\log_{10}$ of average standard deviation vs. $\log_{10}$ of window length. Following [41], we classify a test particle as "random-walk" if the slope of this graph lies between 0.22 and 0.95 and the correlation coefficient is > 0.85. If the correlation coefficient is between 0.7 and 0.85 the dynamical class is found qualitatively. We consider a test particle to be a "resonance-hopping" Centaur if the slope is outside of this range and the correlation coefficient is < 0.9. Otherwise the dynamical class is found qualitatively.

## Results and Discussion

Figure 1 shows the distribution of the test particles studied in this work in semi-major axis, at the start and end of our simulations (i.e. at $t = 0$ and 3 Myr). Of the 8,022 particles studied, 4,967 test particles survive after the full integration time, and move with semi-major axes in the range 5 AU – 35 AU at that time. Comparison of the two histograms shows that many of the test particles have diffused throughout the Solar system, escaping from the vicinity of the 2:1 resonance with Neptune in the process. The shape of the histogram at 3 Myr is roughly Gaussian with a peak within 19 AU – 19.5 AU. This overlaps the initial region and the orbit of Uranus.

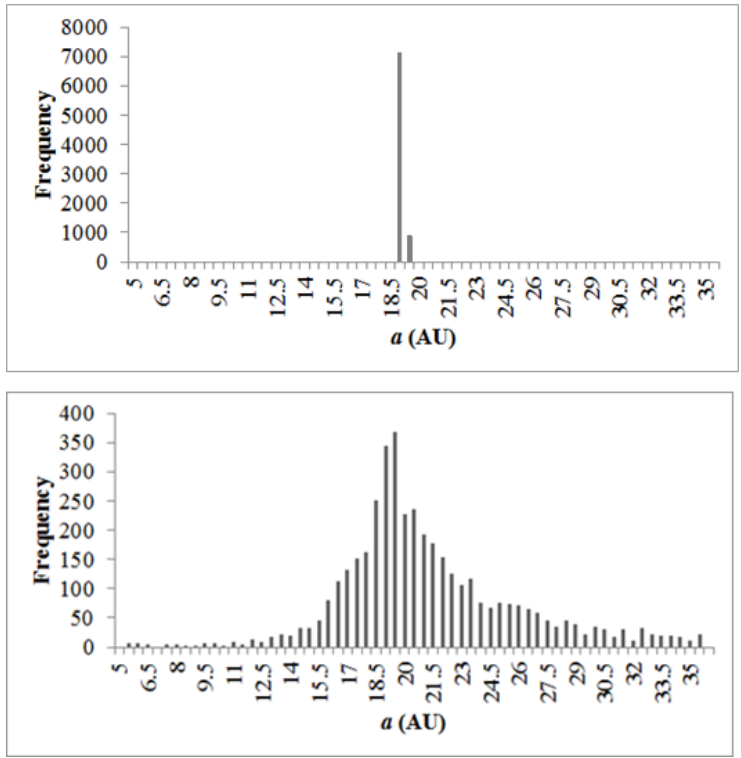

*Figure 1. A histogram of the semi-major axis of test particles at the start (top; t = 0 Myr) and end (bottom; t = 3 Myr) of our simulations. The bin size is 0.5 AU for both histograms. At t = 0 Myr, 7,128 test particles are in the bin between 18.5 AU and 19 AU, and 894 test particles are in the bin between 19 AU and 19.5 AU. At the end of the simulations, 4,967 test particles survive in the range 5 AU < a < 35 AU. The diffusion of the test particles over time is clearly seen.*

The number of test particles entering each class of small body at some point during their lifetime and the average time spent in each class is shown in Table 2. Of the classes inward from the Centaur region, the Jupiter Family Comet class (JFC) had the highest percentage of test particles entering the class, at 35.7%. Only 1.9% of our test particles became Encke-type comets. This result is not particularly unexpected – Encke-type comets are almost decoupled from the influence of Jupiter, and so are relatively dynamically stable. However, that stability in turn means that it is relatively difficult for objects to become captured on such orbits in the first place. 5.7% of the test particles studied became Mars crossers and a mere 3.1% became Earth crossers. Thus, the possibility of a Centaur coming from an orbit near Uranus and colliding with Earth within 3 Myr is small but nonzero. Of the classes outward from the Centaur region, KBO had the highest percentage of test particles entering the class at 53.7% and Ambi-Neptunian Object was 2nd at 47.8%. Only 0.11% became SDOs.

In regards to time spent in each class, the Centaur class had the highest average percentage of occupancy time of any class at 75% of a test particle's lifetime. Earth crossers and Mars crossers had the lowest average percentages of occupancy time overall at 3% and 3.7% respectively. JFCs spent on average 26% of their lives in the class. Encke-type comets had the highest average time of occupancy of the inward classes at 33%. Of the classes outward from the Centaur region KBO had the highest average occupancy time at 14.9% of a test particle's lifetime. The Ambi-Neptunian Object class had an average occupancy time of 27%.

About one-third of the test particles were removed from the integration by entering the Oort Cloud, obtaining a barycentric distance greater than 1,000 AU or by solar system ejection. 1.9% of test particles collided with the Sun.

*Table 2 the number of test particles that entered each class of small body out of 8,022 and the average percentage of their lives spent in that class once there.*

| Class | Oort Cloud | Encke-Type Comet | JFC | Centaur | Halley-Type Comet | KBO | SDO | Earth Crosser |
|---|---|---|---|---|---|---|---|---|
| **TOTAL** | 399 | 154 | 2,864 | 8,022 | 109 | 4,312 | 9 | 249 |
| **Avg %Time** | - | 33% | 26% | 75% | 14% | 14.9% | 10% | 3% |

| Class | Ambi-Neptunian Object | Mars Crosser |
|---|---|---|
| **TOTAL** | 3,831 | 455 |
| **Avg %Time** | 27% | 3.7% |

Qualitative analysis showed that, whilst a given test particle displayed libration-like behavior about the 2:1 resonance, its osculating semi-major axis remained within 1.15 AU of the resonance location, and its mean semi-major axis remained within 0.125 AU of that location while a test particle displayed libration-like behavior about the resonance. Using these bounds, it was discovered that 2,036 test particles librated within the resonance for at least 10 kyr, 97 for 100 kyr or more and 2 for more than 1 Myr. The mean libration time for test particles that were trapped for at least 10 kyr was found to be 27 kyr, and the median time was 10 kyr. The longest libration time observed was 1,100 kyr, as shown in Figure 2.

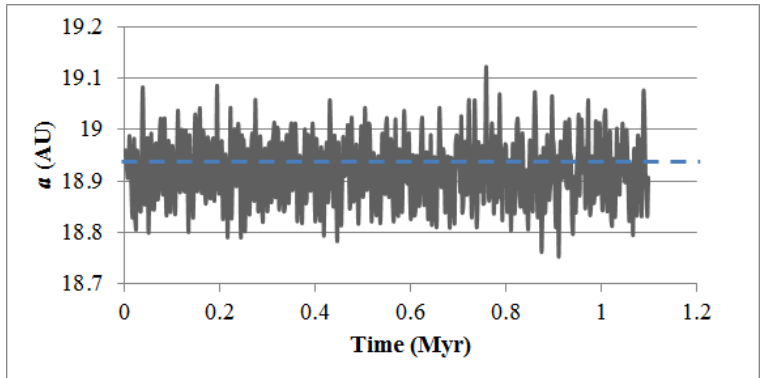

*Figure 2. The libration of the most long lived 2:1 resonant object in our sample. The dashed line denotes the central location of the 2:1 resonance with Neptune. The initial orbital parameters for this test particle were a = 18.95 AU, e = 0.205 and i = 29.2°.*

The dynamical classes of a random sample of 218 Earth-crossing test particles were determined. Given the recent interest in potentially hazardous objects, the particles that became Earth-crossing in this study are of particular interest. 35% of the test particles in the sample were identified as resonance-hopping Centaurs, whilst 65% were found to be random-walk Centaurs. Thus, the most likely method of dynamical transport of a Centaur to the inner Solar system is a random walk. Two case studies are shown in Figures 3 and 4.

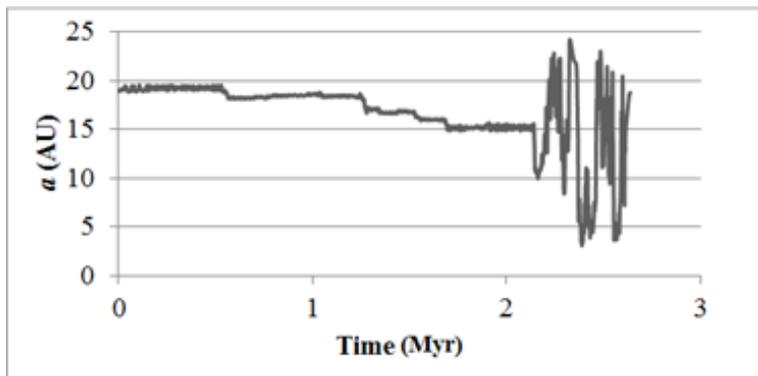

*Figure 3. The first 2 Myr in the evolution of the semi-major axis of an exemplar test particle that eventually became an Earth-crossing object. Throughout this period, the particle is best classified as a resonance-hopping Centaur. Note the periods of libration in various resonances, which appear as nearly horizontal bands in the graph. This test particle later became an Earth crosser at 2.39 Myr. Data points beyond 25 AU are not shown, for clarity. Its initial orbital parameters were a = 18.99 AU, e = 0.063 and i = 37° respectively.*

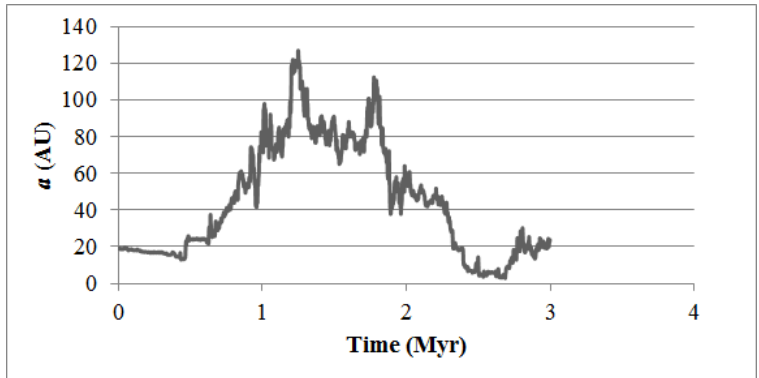

*Figure 4. The evolution of the semi-major axis of an exemplar test particle that evolved to become an Earth-crossing object after a protracted period of random walking throughout the outer Solar system. It is classified as a random-walk Centaur. Note the lack of horizontal bands in the graph (aside from a short resonant capture at ~600 kyr). The test particle became an Earth crosser at about 2.65 Myr. Its initial orbital parameters were a = 19.02 AU, e = 0.44 and i = 23° respectively.*

## Conclusions and Future Work

Dynamical simulations were carried out of the evolution of 8,022 massless test particles, initially located in the vicinity of the 2:1 Neptunian mean-motion resonance, over a period of 3 Myr. The average libration time of test particles displaying libration-like behavior in the vicinity of the resonance for at least 10 kyr was found to be 27 kyr, and the median time was 10 kyr.

Just over one-third of test particles became JFCs, and just over half became KBOs. A negligible amount of test particles became SDOs which shows that most Centaurs do not evolve into orbits beyond the Kuiper belt with perihelia beyond Neptune's orbit. Less than 6% became Mars crossers and 3.1% Earth Crossers. This shows that the odds of a Centaur diffusing from an orbit near Uranus to an Earth-crossing orbit within 3 Myr is low but non-zero. Thus, this study supports the idea that Centaurs do represent a threat to Earth.

Of the taxonomical classes inward from the resonance, the Encke-type comet class had the highest average occupancy time at 33%. Of the outward classes, KBO had the highest average occupancy time at 14.9% of a test particle's lifetime.

65% of a sample of Earth crossers were random walkers and 35% resonance-hoppers. Graphs of semi-major axis vs. time for resonance-hopping Centaurs tend to have long horizontal bands, and those of the random-walk Centaurs tend to lack long horizontal bands.

Future work will be to determine the separatrices (boundaries) of the resonance in *a-e* space and the likelihood of capture as a Trojan of any Jovian planet.